\documentclass[11pt, oneside]{article}   	
\usepackage{geometry}                		
\usepackage{graphicx}				
\usepackage{amssymb}
\usepackage{amsmath}

\usepackage{tikz}
\usetikzlibrary{arrows,scopes}
\newcommand{\rc}{%
\resizebox{!}{1.25ex}{%
    \begin{tikzpicture}[>=round cap]
        \clip (0.09em,-0.05ex) rectangle (0.61em,0.81ex);
        \draw [line width=.11ex, <->, rounded corners=0.13ex] (0.1em,0.1ex) .. controls (0.24em,0.4ex) .. (0.35em,0.8ex) .. controls (0.29em,0.725ex) .. (0.25em,0.6ex) .. controls (0.7em,0.8ex) and (0.08em,-0.4ex) .. (0.55em,0.25ex);
    \end{tikzpicture}%
}%
}
\newcommand{\brc}{%
\resizebox{!}{1.3ex}{%
    \begin{tikzpicture}[>=round cap]
        \clip (0.085em,-0.1ex) rectangle (0.61em,0.875ex);
        \draw [line width=.2ex, <->, rounded corners=0.13ex] (0.1em,0.1ex) .. controls (0.24em,0.4ex) .. (0.35em,0.8ex) .. controls (0.29em,0.725ex) .. (0.25em,0.6ex) .. controls (0.7em,0.8ex) and (0.08em,-0.4ex) .. (0.55em,0.25ex);
    \end{tikzpicture}%
}%
}
\newcommand{\hrc}{\hat{\brc}}
\newcommand\refeq[1]{(\ref{#1})}
\newcommand\reffig[1]{Figure~\ref{#1}}

\newcommand\cites[1]{[\ref{#1}]}
\newcommand\meskip{\, \, \, \, \, \, \, \, \, \, \, }
\newcommand\rad{\hbox{\tiny{rad}}}

\title{The CUBE Virtual Reality Immersion}
\author{L.\ Estridge, J.\ Franklin\\
Physics Department \\ Reed College}
\date{December 1st, 2025}							

\begin{document}
\maketitle

\abstract{The purpose of this note is to introduce the CUBE, a virtual reality immersion that was developed to help visualize electromagnetic fields, particularly the less familiar radiation fields students typically encounter in upper level physics courses.  We discuss the pedagogical motivation for different features found in the software, and provide a brief overview of its use.
}

\section{Introduction}
There are a variety of ways that instructors build intuition about electromagnetic fields.  In a first year physics course, we use electric field lines, literal ``lines of force," to make quantitative predictions about test particle motion, and define the notion of flux.  Once magnetostatics has been introduced, we generally sketch simple magnetic field lines (those around steady, current carrying wires, or inside toroidal solenoids), but omit more involved magnetic field line considerations, which often spoil the immediate utility of the concept (like, for example, the mismatch between field line continuity and the association of field line density with field magnitude~\cites{SLEPIAN}).  Indeed, both electric field lines and magnetic field lines have subtleties that are both well-known and interesting, and have been periodically discussed at the advanced undergraduate level, see~\cites{FGS}~\cites{KZTB} for a recent example.

Once electrodynamics has been introduced, in upper level coursework, instructors often favor contour plots of field magnitudes (or their squares) which are related to the Poynting vector and energy deposition at surfaces.  Such plots can be used to separate out the effects of the static ``Coulomb" field contributions from the radiation field pieces, which helps understand the new geometry of the far field.  But showing contour plots in the classroom typically involves using highly symmetric source motion, and taking slices of fields (electric, magnetic, or the Poynting vector itself) to generate two-dimensional graphs.

Virtual reality (VR) provides a fully three dimensional environment in which to visualize physical phenomena.  This environment has been used in a variety of ways;  an early application was in electromagnetic field design~\cites{CAVE}.  The educational utility of VR has long been clear -- virtual worlds like ScienceSpace~\cites{SSPACE} introduced a suite of tools covering high school physics concepts in VR, and tested the utility of that introduction.  More recently, with the advent of relatively inexpensive, easily programmable virtual reality platforms, the use of three dimensional visualization tools has become more widespread, including applications for lab instruction~\cites{MAROON}, and, most relevant for us, electricity and magnetism (E\&M)~\cites{ESPlay}~\cites{MEVAL}.
The purpose of this paper is to advertise a new addition to the electromagnetic VR visualization library, one that benefits from both hardware and software advances that have occurred over the last five years or so.  

\begin{figure}[htbp] 
   \centering
   \includegraphics[width=2in]{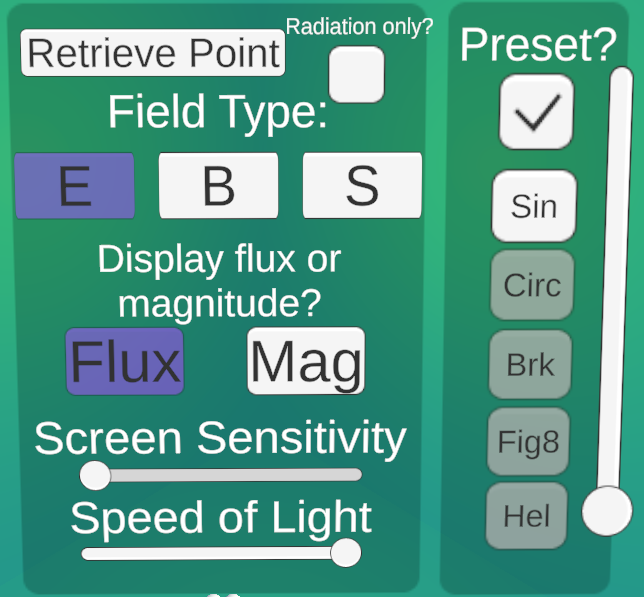} 
   \caption{The menu where users select which type of field, ${\bf E}$, ${\bf B}$, or ${\bf S}$ (the Poynting vector) to display, whether to show the magnitude ($E \equiv |\bf E|$) or the flux (${\bf E} \cdot d{\bf a}$) at the wall's surfaces, and other display options.  }
   \label{fig:menu}
\end{figure}

The CUBE software~\cites{LET}
 described here was produced using the Unity environment, which presents a relatively low bar to VR application development entry.  The (C)ube (U)bserver of (B)magnetic and (E)lectric fields (in a nod to the CAVE), is able to do much more, and much more quickly, than previous, similar software~\cites{RYDER}, using GPU acceleration via shaders. The CUBE demonstration places the user within a cubical room with walls that display contour maps representing the electromagnetic fields generated by a point charge. The user is able to grab the point charge, represented by a small sphere, and move it to observe how different types of motion affect the generated fields. A menu, seen in~\reffig{fig:menu}, is attached to the left handset controller and allows the user to select display options, such as:
\begin{itemize}
   \item Observe ${\bf E}$ (electric field), ${\bf B}$ (magnetic field), or ${\bf S}$ (the Poynting vector).
   \item Display either the magnitudes of the field vectors at the surface of the walls or the infinitesimal flux through the walls (e.g.\ ${\bf E} \cdot d{\bf a}$).
   \item Change the speed of light to observe relativistic effects. 
   \item Only display the radiation field.
   \item Set the point charge to move automatically along one of several pre-defined paths.
\end{itemize}

Our goals in producing The CUBE were to:  1.\ provide a visualization scheme for the ``near" field (the portion of the electric and magnetic fields that fall off like $1/r^2$) that highlights its geometry, 2.\  allow students to ``see" the radiation field separate from the rest of ${\bf E}$ and ${\bf B}$, to build intuition about its behavior, and 3.\  demonstrate the difference between non-relativistic and relativistic source motion, particularly as measured by the energy flowing through a surface.  In each of the next three sections, we will detail our work on these three targets. We conclude with some thoughts on low-hanging fruit for additional features, and a focused survey of implementation ideas.

\section{Static Fields}

\begin{figure}[htbp] 
   \centering
   \includegraphics[width=2in]{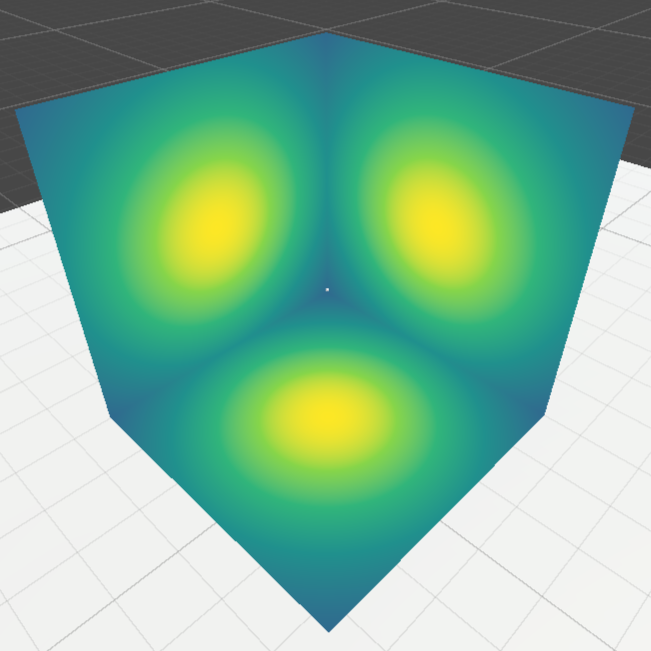} 
   \caption{A screenshot, from outside the CUBE in the Unity editor, showing the magnitude of the electric field on the walls for a charge located at the center of the CUBE.}
   \label{fig:staticcharge}
\end{figure}

Our first task was the visualization of fields for static charges and steady currents.  For a point charge fixed at the center (origin) of the cubical ``room," the electric field is ${\bf E} \sim \hat{\bf r}/r^2$, and we can display the field magnitude at the walls using a heat map, with lighter colors associated with larger magnitudes.  One gets the familiar diffuse circle at the center of each wall, shown, for reference in~\reffig{fig:staticcharge}.

The (approximate, in the slow motion limit) electric and magnetic fields are, relative to the origin at the center,
\begin{equation}
{\bf E} = \frac{q}{4 \pi \epsilon_0} \frac{\brc}{\rc^{ 3}} \meskip {\bf B} = \frac{\mu_0 q {\bf v} \times \brc}{4 \pi\rc^{ 3}}, 
\end{equation}
where $\brc \equiv {\bf r} - {\bf r}'$ is the vector pointing from the charge (at ${\bf r}'$) to locations on the walls (at ${\bf r}$), and ${\bf v}$ is the velocity of the particle as reported by Unity.  For completeness, and later use, we also compute the Poynting vector, ${\bf S} = {\bf E}\times {\bf B}/\mu_0$.

Since the fields are never exactly zero within the room and its walls, the contour plot scales the values of the field magnitudes when associating heat map ``colors," so overall coefficients are ignored, i.e.\ the contour plot is produced from the relative values of ${\brc}/\rc^{ 3}$ and ${\bf v} \times {\brc}/\rc^{ 3}$.  Distances in Unity are given in meters, and speeds are quoted in meters per second.

\begin{figure}[htbp] 
   \centering
   \includegraphics[width=2in]{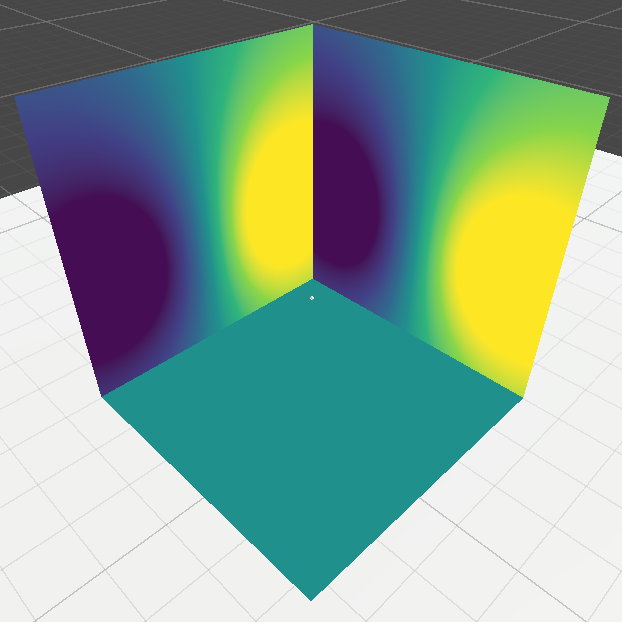} 
   \hspace{0.5cm}
   \includegraphics[width=2in]{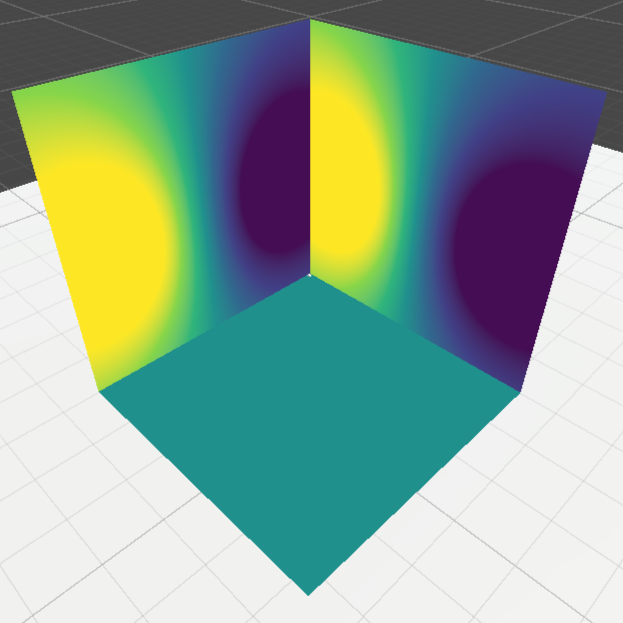} 
   \caption{On the left, a screenshot of the flux of ${\bf B}$ at the walls of the cube, for a charge moving down.  On the right, the same screenshot for a charge moving up.}
   \label{fig:BdotA}
\end{figure}

We wanted a way for users to associate the magnetic field's direction with the direction of its source's motion, an association that is not captured by the field magnitude.  So we introduced a mode in which the infinitesimal flux through the walls, ${\bf E} \cdot d{\bf a}$ and ${\bf B} \cdot d{\bf a}$, can be displayed, again using a scaled contour plot on the wall surfaces.   Light colors indicate that the field vectors are oriented into the wall, with dark colors indicating they are oriented out of the wall.  For the magnetic flux, the pattern of light and dark flips as users move the charge up or down, as shown in~\reffig{fig:BdotA}.

It is worth noting that the cubical geometry of the CUBE is useful in visualizing the field information -- some users place the charge near the corners of the CUBE to see how the field is projected onto three surfaces at once to get a better sense of the three dimensional structure of the electromagnetic fields and Poynting vector.  This would not be possible using a spherical ``room," although that geometry has other benefits that recommend it.

\section{Dynamic Fields}

Moving on to the full electric and magnetic fields of a point charge moving along a prescribed trajectory ${\bf w}(t)$, from~\cites{DJG}, 
\begin{equation} \label{EB}
{\bf E}({\bf r}, t) = \frac{q}{4 \pi \epsilon_0} \frac{\rc}{({\bf u} \cdot \brc)^3}\left[ (c^2 - v^2) {\bf u} + \brc \times ({\bf u} \times {\bf a})\right]
\meskip {\bf B}({\bf r},t) = \frac{1}{c} \hrc \times {\bf E}({\bf r},t),
\end{equation}
with $\brc \equiv {\bf r} - {\bf w}(t_r)$, the vector pointing from the retarded time location of the charge, ${\bf w}(t_r)$, to the field point ${\bf r}$ at time $t$, where the retarded time is defined implicitly by: $c (t - t_r) = | {\bf r} - {\bf w}(t_r)|$.  The velocity and acceleration vectors associated with the charge are also evaluated at the retarded time, ${\bf v} \equiv \dot{\bf w}(t_r)$, and ${\bf a} \equiv \ddot{\bf w}(t_r)$.  The auxiliary vector, ${\bf u}$, is defined to be: ${\bf u} \equiv c \hrc - {\bf v}$.  For now, the visual ``feel" of the VR room is on the order of meters, so we do not expect the retarded time to be significantly different from the coordinate time at the field point (on the walls) as we move a charge around in the room.  We therefore omit the retarded time piece of the calculation in what follows -- that is, we set $t_r = t$.

Unlike the velocity vector, which is a measured quantity, the acceleration needs to be calculated.  The velocity data is collected with a temporal cadence of $\Delta t = 0.02$ s.  We know the measured (and interpreted through Unity) velocity at times $t$ and $t + \Delta t$, and use the rough approximation:
\begin{equation}
{\bf a}(t) \approx \frac{{\bf v}(t + \Delta t) - {\bf v}(t)}{\Delta t}
\end{equation}
for the acceleration at time $t$, in~\refeq{EB}.

For a human moving a charge around using the VR handsets, the dominant piece of the electric and magnetic fields will come from the $1/\rc^{ 2}$ term in ${\bf E}$ from~\refeq{EB} (the first term in parentheses, with $v \rightarrow 0$).  That's fine, but we'd also like to be able to see the radiation piece of the fields, defined by:
\begin{equation}\label{EBrad}
{\bf E}_{\rad}({\bf r}, t) = \frac{q}{4 \pi \epsilon_0} \frac{\rc}{({\bf u} \cdot \brc)^3}\left[ \brc \times ({\bf u} \times {\bf a})\right]
\meskip {\bf B}_{\rad}({\bf r},t) = \frac{1}{c} \hrc \times {\bf E}_{\rad}({\bf r},t),
\end{equation}
and within the CUBE's menu area, there is a checkbox that allows us to select just this radiation ($1/\rc$) component of the full field, the portion that carries energy out to infinity.  As in the static case, we can choose to display the fields or the Poynting vector (either full, or just radiation), and we can again choose to project either the magnitude or flux of these vectors onto the walls.

\begin{figure}[htbp] 
   \centering
   \includegraphics[width=1.75in]{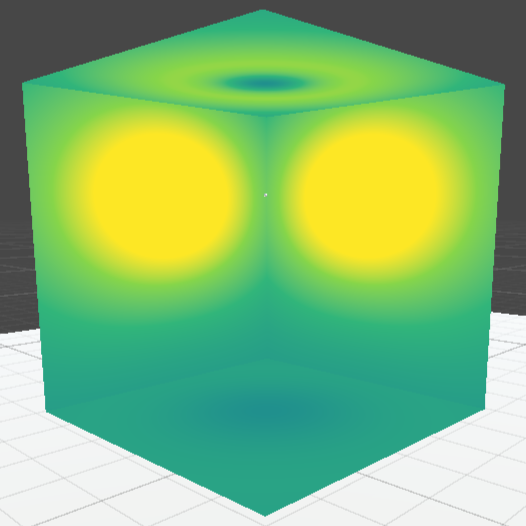}
   \hspace{0.1cm} 
   \includegraphics[width=1.75in]{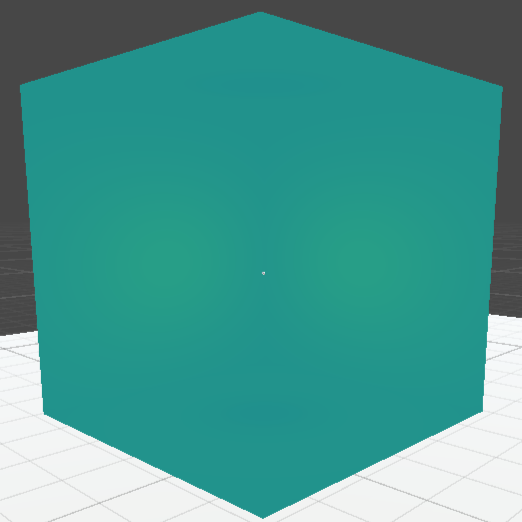}
   \hspace{0.1cm}
   \includegraphics[width=1.75in]{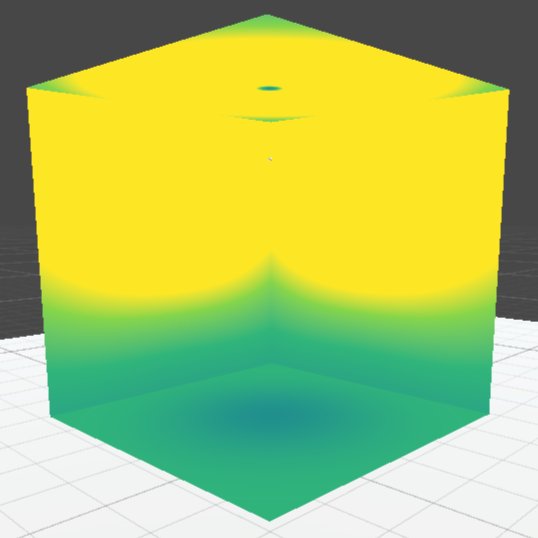}
   \caption{On the left is a screenshot of four walls of the CUBE as a user starts moving a charge vertically  in an oscillatory manner -- you can see large radiated power deposited on the walls (where $\sin\theta \approx 1$ from~\refeq{LarmorS}), with little on the ceiling and floors.  In the middle image, the same setup shown as the charge moves through an ``equilibrium" position at the center of the cube, now the Poynting vector is small everywhere because ${\bf a} \sim 0$.  Finally, on the right, we see the Poynting vector flux as the charge nears the top of its trajectory, where acceleration is again large (slowing to a stop).}
   \label{fig:LarmorS}
\end{figure}

We imagine that the primary utility of the full relativistic solution to Maxwell's equations for a charge shown in~\refeq{EB} will be in building intuition about the new, radiation piece, of the fields.  In particular, the radiated power can be surprising.  For the nonrelativistic limit, the Poynting vector associated with the Larmor power has the form
\begin{equation}\label{LarmorS}
{\bf S} = \frac{\mu_0 q^2 a^2}{16 \pi^2 c} \frac{\sin^2\theta}{\rc^{ 2}} \hrc,
\end{equation}
where $\theta$ is the angle between the field point and the acceleration of the charge (to be clear, we do not use this form directly, it is the low speed limit of ${\bf S}_{\rad} \equiv {\bf E}_{\rad} \times {\bf B}_{\rad}/\mu_0$ that we compute from the radiation fields).  The Poynting vector is largest in magnitude at points perpendicular to the charge's instantaneous acceleration.  If a user moves a charge up and down rhythmically in the CUBE, and displays the Poynting flux on the walls, they will see large values on the side walls, with small values directly above and below the charge on the ceiling and floor.  An example of this type of motion, and the Poynting flux on four walls, is shown in~\reffig{fig:LarmorS}.

\section{Relativistic Motion}
While the energy distribution of the radiation fields at slow speeds is for the most part unfamiliar to students entering their upper level physics courses, even less familiar is the case of sources moving at relativistic speeds.  In order to evaluate the relativistic Poynting vector, we have introduced a ``speed of light" slider (see~\reffig{fig:menu}) that lets the user set $c$ to be small, commensurate with speeds at which a person could wave their arms.  At relativistic source speeds, the geometry of the resulting radiated power profile is changed as compared with, for example~\refeq{LarmorS}.  Now the power is concentrated in the ``forward" (as determined by ${\bf v}$) direction, although it still has a hole directly in front of and behind the particle's acceleration vector.~\footnote{It is important to note that ${\bf S}_{\rad}$ does {\it not} lead to Li\'enard's generalization of the Larmor power, that power is measured with respect to the retarded time, $t_r$, not the coordinate time at the field point, which is what ${\bf S}_{\rad} \cdot d{\bf a}$ gives.  Nevertheless, they share the same qualitative behavior.}  Users can grab a charge and move it around, but can now do so with relativistic speeds, and both see the relativistic power pattern, and compare it with the same pattern for a large speed of light (the non-relativistic case).

\begin{figure}[htbp] 
   \centering
   \includegraphics[width=2in]{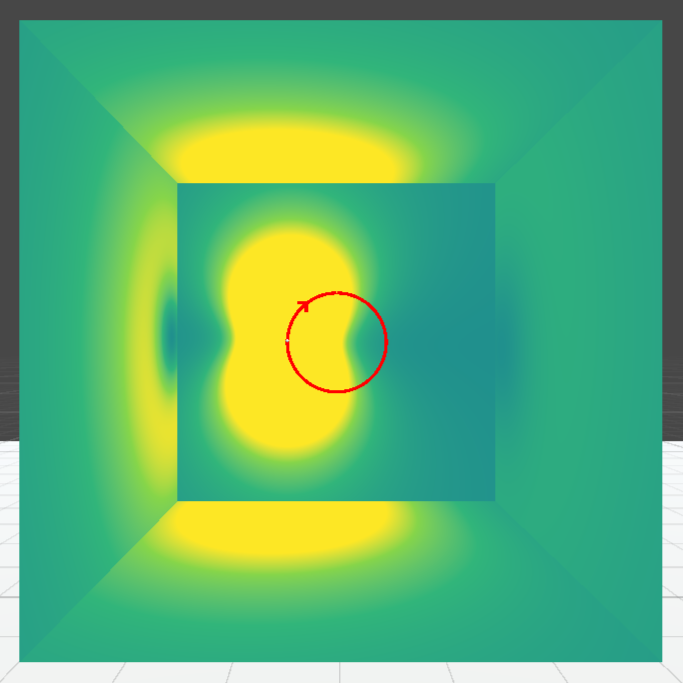}
   \hspace{0.5cm}
   \includegraphics[width=2in]{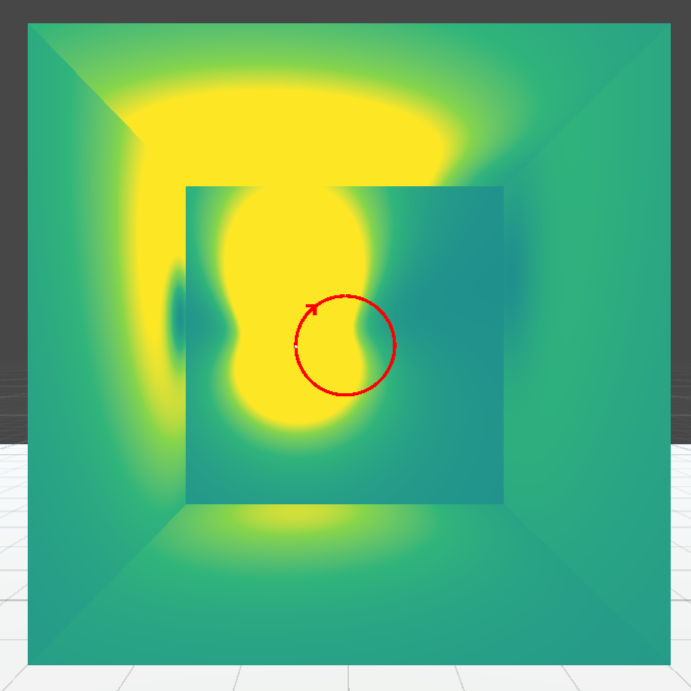}
   \caption{Slow uniform circular motion, on the left, has power that is perpendicular to the acceleration, and equal in magnitude in both the forward and backward direction.  For a particle moving close to the speed of light (right), the power is focused in the forward direction.}
   \label{fig:ucirc}
\end{figure}

We have provided some ``pre-loaded" trajectories to make it easier for students to focus on the radiated power (as opposed to waving the controllers around attempting to approximate sinusoidal oscillation, for example), and these can be selected from the CUBE's menu.  It is easy to see the difference between the non-relativistic and relativistic power for a charge moving in uniform circular motion, one of the selectable trajectories.  In~\reffig{fig:ucirc}, we can see the Poynting flux on the walls for a charge moving in a circle at low speeds (left), where the dominant power is in front of and behind the particle (perpendicular to its inward-directed acceleration, as predicted by~\refeq{LarmorS}), and for speeds near the speed of light (right), where the power is focused in the forward direction.

\section{Conclusion and Future Work}

We have provided a brief overview of the CUBE, with a focus on the pedagogical motivation of some of its features, in particular:
\begin{itemize}
\item Its ability to display magnitude or flux for the three most relevant vector fields: ${\bf E}$, ${\bf B}$, and ${\bf S} = {\bf E} \times {\bf B}/\mu_0$, for a particle moving along an arbitrary, user- or pre-defined trajectory.
\item Its cubical geometry, allowing for a fully three-dimensional investigation of the fields of a moving charge.
\item Its focus on radiation, both through calculation of the ``radiation-only" fields and a movable speed of light to probe both radiation generated by non-relativistic sources and the less familiar relativistic sources.
\end{itemize}

This note is only meant to advertise the basic features of the CUBE, we invite you to try it out for yourself.  The CUBE, its source code, with full documentation and developer information is available at: \texttt{https://github.com/ReedPhysicsVR/TheCUBE}.  We hope that people download, use, and extend this work.  A lot of effort has gone into making the code readable, and the portions of Unity that are used are carefully documented, providing a path to self-study for folks unfamiliar with this powerful three-dimensional development environment.  There are a few immediate avenues of interest to us, things we hope to implement in the near future that might provide starting points for other developers, including:
\begin{itemize}
\item  A spherical CUBE:  We'd like to make a version in which the ``room" has surface that is spherical, matching the direction of the radiation Poynting vector for a charge moving near the center of the room.  The sphere's radius could be changed to probe the near and far fields.
\item The CUBE has size that does not require the retarded time evaluation of the source trajectory.  With a variably-sized spherical shell, and a changeable speed of light, this will not always be the case, so we're hoping to add a full retarded time calculation for both pre-set trajectories and, in some form, user-specified trajectories.
\item By solving the first order differential equation
\begin{equation}\label{FODE}
\frac{d {\bf r}(u)}{du} = {\bf E}({\bf r}(u))
\end{equation}
for a given electric field ${\bf E}$, we can produce a field line provided the initial condition (starting point of the line) ${\bf r}_0$.  We'd like to allow users to select points in space (in real time), solve~\refeq{FODE} with those points as initial conditions, and connect those points to visualize electric, magnetic, or Poynting vector field lines.
\end{itemize}

\section{Appendix:  Implementation Notes}

In this section, we include a couple of comments on the implementation of the CUBE -- these are technical notes that motivate choices that were made in developing the software, so we felt they would be appropriate here, rather than in the documentation of the code or instructional information available from the website.

\subsection{Scaling the Speed of Light}
\begin{figure}[htbp] 
   \centering
   \includegraphics[width=2in]{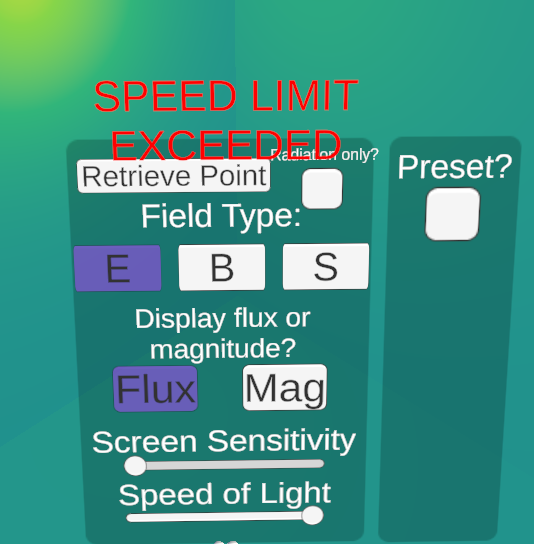} 
   \caption{When the particle exceeds the simulation's set speed of light, this text is displayed on the settings menu.}
   \label{fig:speedlimit}
\end{figure}

The ability to change the speed of light in the simulation is incredibly useful for displaying the relativistic effects, however it does present some challenges in terms of implementation. The goal is to the lower the speed of light to a point where the relativistic effects become clear (and sometimes extreme) without allowing the particle to move faster than the speed of light. To attempt to address this, the speed of light slider as displayed in~\reffig{fig:menu} does not scale the speed of light linearly, but on an exponential curve. This allows the user to ``zoom'' past the values for speed of light where the relativistic effects are not terribly visible to the naked eye to a region of values which convey the relativistic effects well while still mostly sitting above usual speeds the particle can travel at. It is important to note that the particle's velocity is not capped at this time and can, at low speeds of light and high particle velocities, exceed the speed of light. When this happens, the menu displays a ``speed limit'' sign (seen in \reffig{fig:speedlimit}) to warn the user that the field's appearance may no longer be accurate.

\subsection{Choosing Colors}
\begin{figure}[htbp] 
   \centering
   \includegraphics[width=1.75in]{StaticBUpZoomOut.PNG}
   \hspace{0.1cm} 
   \includegraphics[width=1.75in]{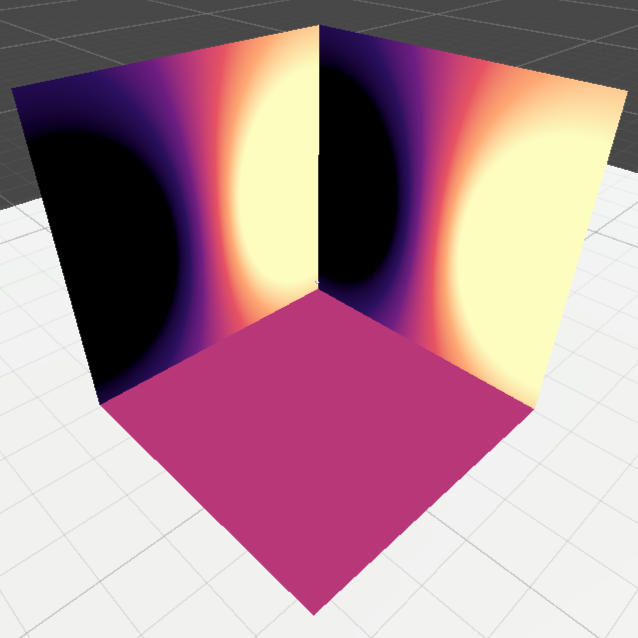}
   \hspace{0.1cm}
   \includegraphics[width=1.75in]{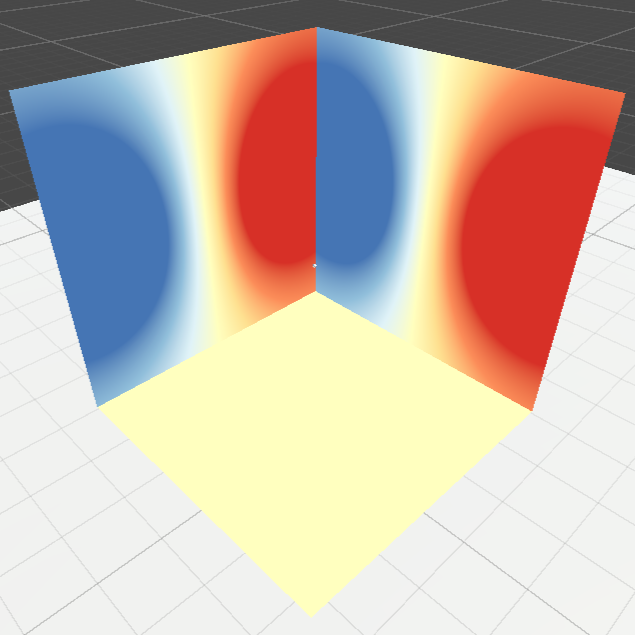} 
   \caption{The three color palette options: viridis (left), magma (middle), and diverging (right). The heatmaps are displaying the static magnetic field's flux through the walls as the particle moves downwards.}
   \label{fig:colorpalletes}
\end{figure}

Heat maps offer an aesthetically pleasing way to represent data, but using color to convey information can be problematic if one is not careful and intentional with what colors are chosen. Human perception of color is not linear and can vary with different types of colorblindness, so a color gradient must be chosen such that the heat maps accurately portray the vector field characteristics to the largest population possible. Three color gradients were chosen: viridis, magma, and a diverging red-blue color palette, all displayed in~\reffig{fig:colorpalletes}. Viridis and magma are both colorblind-friendly and were specifically developed for the accurate portrayal of scientific data regardless of colorblindness~\cites{VIRIDIS}. The diverging red-blue color palette is not colorblind-friendly, however it is still included for those who can perceive all colors since it is the most clear in portraying the split between positive and negative values for the flux of the field through the walls. Taking all these points into account, viridis was chosen as the default color palette, and the user is able to toggle through the three palettes to choose the one that is most comfortable and perceptible for themselves.

\end{document}